\documentclass[comment, prl,twocolumn,nofootinbib, preprintnumbers, superscriptaddress]{revtex4}

\usepackage{amsmath,amssymb,slashed,braket}
\usepackage{graphicx}
\usepackage{epstopdf}
\usepackage{float}
\usepackage{cancel}
\usepackage[colorlinks=true,
            linkcolor=blue,
            urlcolor=blue,
            citecolor=green,          
            bookmarks=true,
            bookmarksnumbered=true,
            breaklinks=true,
            pdfpagemode=Fullscreen,
            pdfstartview=FitBH]{hyperref}

\usepackage[normalem]{ulem}
\usepackage{color}

\definecolor{Orange}{cmyk}{0,0.61,0.87,0}
\definecolor{JungleGreen}{cmyk}{0.99,0,0.52,0}
\definecolor{OliveGreen}{cmyk}{0.64,0,0.95,0.40}
\definecolor{Brown}{cmyk}{0,0.81,1,0.60}
\definecolor{RoyalBlue}{cmyk}{0.71,0.53,0,0.12}
\definecolor{Gray}{cmyk}{0,0,0,0.40}
\definecolor{LightPink}{cmyk}{0.0,0.25,0,0}
\definecolor{LLightPink}{cmyk}{0.0,0.10,0,0}
\definecolor{LightBlue}{cmyk}{0.25,0,0,0}
\definecolor{LightGray}{cmyk}{0,0,0,0.2}

\usepackage{xcolor}
\definecolor{gesfpurple}{rgb}{0.47,0.19,0.42}

\definecolor{gesflanse}{rgb}{0.00,0.50,0.50}

\definecolor{gesfblue}{rgb}{0.08,0.42,0.76}

\definecolor{gesfred}{rgb}{1,0,0}

\definecolor{gesfwhite}{rgb}{1,1,1}

\definecolor{gesfblack}{rgb}{0,0,0}

\newcommand{\geqn}[1]{Eq.\,\hypersetup{linkcolor=blue}(\ref{#1})\hypersetup{linkcolor=blue}}
\newcommand{\gfig}[1]{{\hypersetup{linkcolor=violet}Fig.\,\ref{#1}\hypersetup{linkcolor=blue}}}

\graphicspath{{figs/}}

\begin{document}

\title{
Testing the RG Running of the Leptonic Dirac CP Phase with Reactor Neutrinos
}

\author{Shao-Feng Ge}
\email{gesf@sjtu.edu.cn}
\affiliation{State Key Laboratory of Dark Matter Physics, Tsung-Dao Lee Institute \& School of Physics and Astronomy, Shanghai Jiao Tong University, China}
\affiliation{Key Laboratory for Particle Astrophysics and Cosmology (MOE) \& Shanghai Key Laboratory for Particle Physics and Cosmology, Shanghai Jiao Tong University, Shanghai 200240, China}

\author{Chui-Fan Kong}
\email{kongcf@sjtu.edu.cn}
\affiliation{State Key Laboratory of Dark Matter Physics, Tsung-Dao Lee Institute \& School of Physics and Astronomy, Shanghai Jiao Tong University, China}
\affiliation{Key Laboratory for Particle Astrophysics and Cosmology (MOE) \& Shanghai Key Laboratory for Particle Physics and Cosmology, Shanghai Jiao Tong University, Shanghai 200240, China}

\author{Pedro Pasquini}
\email{pedrosimpas@g.ecc.u-tokyo.ac.jp}
\affiliation{Department of Physics, University of Tokyo, Bunkyo-ku, Tokyo 113-0033, Japan}
\affiliation{Instituto de Física Gleb Wataghin - Universidade Estadual de Campinas, 13083-859, Campinas SP, Brazil}
\begin{abstract}
We propose the possibility of using the near detector at
reactor neutrino experiments to probe the renormalization group 
(RG) running effect on the leptonic Dirac CP phase $\delta_D$. 
Although the reactor neutrino oscillation cannot directly measure
$\delta_D$, it can probe the deviation 
$\Delta \delta \equiv \delta_D(Q^2_d) - \delta_D(Q^2_p)$ 
caused by the RG running. Being a key element, the mismatched
momentum transfers at neutrino production ($Q^2_p$) and
detection ($Q^2_d$) processes can differ by two orders.
We illustrate this concept with the upcoming Taishan
Antineutrino Observatory (TAO, also known as JUNO-TAO)
experiment and obtain the projected sensitivity to the
CP RG running beta function $\beta_\delta$.
\end{abstract}
 
\maketitle

\textbf{Introduction} --
The charge-parity (CP) symmetry violation (CPV)
is a key element for explaining the 
matter-antimatter asymmetry in our Universe
\cite{Canetti:2012zc,Garbrecht:2018mrp}. In the leptogenesis 
scenarios \cite{Fukugita:1986hr, Buchmuller:2005eh, Davidson:2008bu,DiBari:2012fz,Blanchet:2012bk,Xing:2020ald}, the CPV comes
from the neutrino sector \cite{Branco:2011zb}.
Consequently, measuring the leptonic CP phase
$\delta_D$ at neutrino oscillation experiments
\cite{PDG22-NuCP} provides 
an indirect test of one very important
element of leptogenesis. Confirming
a nonzero and even maximal CP phase is one of
those most important physical goals of the
neutrino oscillation experiments nowadays.

Typically, the leptonic Dirac CP phase $\delta_D$ is
measured by the accelerator-based neutrino experiments.
There is already around $3\,\sigma$ indication
of a maximal leptonic CP phase $\delta_D \sim -90^\circ$
from T2K \cite{T2K:2023smv,T2K24} and NO$\nu$A
\cite{NOvA:2021nfi, NOVA24}
with the $\nu_\mu\rightarrow\nu_e$ and 
$\bar\nu_\mu\rightarrow\bar\nu_e$ appearance 
channels. Although there is some tension,
their combination \cite{joint-fit}
seems quite promising. The next-generation T2HK
\cite{Hyper-Kamiokande:2018ofw} and DUNE
\cite{DUNE:2015lol} will try to confirm a
nonzero $\delta_D$ with more than $5\,\sigma$
sensitivity. Their combination with low-energy
neutrinos from muon decay at rest, such as
TNT2K \cite{Evslin:2015pya} and $\mu$THEIA \cite{Ge:2022iac},
can further narrow the CP phase uncertainty
\cite{Ge:2017qqv, Ge:Proceedings}
and exclude multiple theoretical uncertainties
\cite{Ge:2016xya, Ge:2016dlx}.

With energy below 10\,MeV, the reactor
neutrino oscillation experiments can only measure 
the electron anti-neutrino disappearance
($\bar\nu_e\rightarrow\bar\nu_e$) channel.
It is a common belief that reactor neutrino
experiments alone can not probe the leptonic CP effect
at all. However, this is not necessarily true
in the presence of new physics beyond the
Standard Model (BSM) such as non-unitary mixing 
\cite{Li:2018jgd}, as well as scalar
\cite{Ge:2018uhz} and dark
\cite{Ge:2019tdi, Ge:Proceedings} non-standard interactions,
although not emphasized therein.
In this paper, we provide one more example with
the renormalization group (RG) running of
the leptonic Dirac CP phase $\delta_D$.

In the presence of BSM, the neutrino mixing angles
and Dirac CP phase $\delta_D$ can experience
RG running
\cite{Casas:1999tg,Antusch:2003kp,Antusch:2005gp,Ray:2010rz,Xing:2006sp,Luo:2012ce,Ohlsson:2012pg,Ohlsson:2013xva,Xing:2017mkx,Huang:2018wqp,Babu:2021cxe}.
If such BSM particle is light, the running may be
observed in oscillation experiments as effective
non-unitary mixing. This is because the neutrino
production and detection processes can have different
momentum transfer $Q^2_p$ and $Q^2_d$, respectively.
As a consequence, there is a mismatch between the
two neutrino mixing matrices $U(Q^2_p)$ and $U(Q^2_d)$
at the detection and production vertices 
\cite{Bustamante:2010bf,Babu:2021cxe,Babu:2022non,Ge:2023azz}.
Although both $U(Q^2_p)$ and $U(Q^2_d)$ are unitary
matrices, the non-unitary feature
$U^\dagger(Q^2_p) U(Q^2_d) \neq \mathbb I$ appears
in the neutrino oscillation process.
This  phenomenon can disturb our interpretation of 
the Dirac CP phase $\delta_D$ from the data at
long-baseline accelerator neutrino experiments
\cite{Babu:2021cxe,Ge:2023azz}. It is of ultimate
importance to use experimental data to measure
or constrain such RG running effect to guarantee
the CP phase measurement.

Fortunately, the effect of $\delta_D$ RG runninng
can already appear at the zero-distance limit
\cite{Babu:2021cxe,Ge:2023azz} due to the effective
non-unitarity as explained above. Note that this
effect does not depend on the absolute 
value of the CP phase. It is then possible to use 
the existing data from the high-energy
$E_\nu = \mathcal O(1 \sim 100)\,{\rm GeV}$ 
short-baseline neutrino experiments
\cite{Babu:2021cxe,Ge:2023azz} to put some constraints.
However, the momentum transfer of these high energy
experiments is at least $\mathcal O(100)$\,MeV
as limited by the neutrino production process.
In this paper, we propose that the low-energy
short-baseline reactor experiment has the
advantage to probe thedown to $\mathcal O(1)$\,MeV. We illustrate with
the Taishan Antineutrino Observatory (TAO, also known as JUNO-TAO) experiment \cite{JUNO:2020ijm}
that will start taking data soon.

\vspace{2mm}
\textbf{RG Running and Zero-Distance Effect}
--
In the presence of RG running, the Dirac
CP phase $\delta_D$ can possess scale
dependence parametrized by its $\beta$ function \cite{Antusch:2003kp,Ray:2010rz}, as follows:
\begin{align}
    \frac{d\delta_D}{d\ln \mu^2}
\equiv
  \beta_\delta,
\end{align}
where $\mu^2$ is the renormalization scale.
A widely used choice for $\mu^2$
is the Lorentz-invariant
momentum transfer $|Q^2|$
\cite{Bustamante:2010bf,Babu:2021cxe,Babu:2022non}
that is known as the Gell Mann-Low
scheme \cite{Gell-Mann:1954yli, Wu:2013ei}.
Notice that in general the mixing angles would
enter $\beta_\delta$.
However, the neutrino mixing parameters have already
been precisely measured at the percentage level 
\cite{deSalas:2020pgw,Capozzi:2021fjo,Esteban:2024eli}.
Although these global fits are done without considering
the RG running
effects, a reasonable estimation is that the RG effect
should be within the corresponding error bars. Consequently,
the RG effect of mixing angles and the induced
variation of $\beta_\delta$ should not exceed 10\% as
a rough estimation. Moreover, we are still trying to 
obtain an upper bound of $\beta_\delta$ instead of its 
precise measurement in \cite{Babu:2021cxe,Ge:2023azz}
and our current study. A 10\% variation
from the theoretical side is much smaller than the
experimental uncertainty. It is safe to neglect the
at most 10\% theoretical uncertainty in this first
try of exploring the major phenomenological features
of CP RG running. So we take $\beta_\delta$
and the other mixing parameters as constants to
illustrate the idea which is also consistent with
the BP1 model of \cite{Babu:2022non,Babu:2021cxe}.

For small $\beta_\delta$, the evolution of 
$\delta_D$ can be expanded as power series 
of $\ln |Q^2|$ \cite{Ge:2023azz},
\begin{align}
  \delta_D(Q^2)
& =
  \delta_D(Q^2_0)
+ \beta_\delta 
  \ln\left(\left|\frac{Q^2}{Q^2_0}\right|\right)
+\dots
\label{eq:delta-run}
\end{align}
The reference scale corresponding to new physics is denoted by $Q^2_0$. Then the above evolution of $\delta_D$ in \geqn{eq:delta-run} only apply for $|Q^2| > |Q^2_0|$ and reduces to
a constant value for $|Q^2| < |Q^2_0|$. 
The higher-order terms are typically unimportant for the momentum transfers in the $\mathcal O(1 - 100)$\,MeV$^2$ scale.

For reactor neutrino experiments, the interactions for the
neutrino production and detection processes are different.
More precisely, the momentum transfer $Q^2_p$ of the
production process
can differ from its detection counterpart $Q^2_d$.
According to \geqn{eq:delta-run}, the neutrino mixing 
matrices in the production and detection processes mismatch
with each other. 
Then the neutrino oscillation probability with the 
RG running effect deviates from the conventional one without RG. 
The general expression of the $\nu_\alpha\rightarrow \nu_\beta$ 
vacuum oscillation amplitude is \cite{Babu:2021cxe,Ge:2023azz}
$
  \mathcal{A}_{\beta \alpha}
\equiv 
  \sum_i U_{\beta i}(Q^2_d)
  e^{-iLm_i^2/(2E_\nu)} U^*_{\alpha i}(Q^2_p),
$
which is true for low-energy reactor neutrinos since 
the matter effect can be neglected \cite{Li:2016txk}.
The oscillation amplitude $\mathcal A_{\beta \alpha}$
consists of the neutrino energy $E_\nu$, the neutrino 
mass squared $m^2_i$,
and the propagation distance $L$.
The non-zero flavor transition 
effect occurs in the zero-distance 
limit when the propagation distance 
is small, $L \ll 2 E_\nu / \delta m^2_{ij}$
with $\delta m^2_{ij} \equiv m^2_i - m^2_j$,
but is still macroscopic, $L \gg 1/E_\nu$. 
As a concrete example, the survival probability
for $\nu_e\rightarrow \nu_e$ in the zero-distance
limit is,
\begin{align}
  P_{ee} (Q^2_{d,p}) 
=
  1
- \sin^2\left(\frac{\Delta \delta_D}{2}\right)\sin^2 2\theta_{13},
\label{eq:Pee}
\end{align}
where the difference in CP phase between the detection 
and the production processes
is defined as 
$\Delta \delta_D (Q^2_{d,p}) \equiv \delta_D (Q^2_d)  - \delta_D (Q^2_p)$.
For a small RG running effect, $\beta_\delta \ll 1$,
the transition probability expands to
\begin{align}
  1 - P_{\rm ee}(Q^2_{d,p})
\approx
\left[
  \frac 1 2
  \ln \left( \left| \frac {Q^2_d}{Q^2_p} \right| \right)
  \sin 2 \theta_{13} \beta_\delta
\right]^2.
\label{eq:Pee-expanded}
\end{align}
Since the reactor angle $\theta_{13} \approx 9^\circ$
\cite{deSalas:2020pgw}, the mixing angle prefactor gives
$\sin 2 \theta_{13} / 2 \approx 0.15$.

The above result shows that the zero-distance
effect is a function of the CP phase difference
$\Delta \delta_D$ from the RG running, but not
the absolute value of the Dirac CP
phase $\delta_D$ itself. The survival probability 
goes to unity when no RG effect is present as expected.
Hence the RG running of the Dirac CP phase can
be tested in the upcoming JUNO-TAO experiment
with large statistics. 
More importantly, there is no need to worry about the 
current large uncertainty on the Dirac CP phase.

\vspace{2mm}
\textbf{Mismatched Momentum Transfers}
--
At JUNO-TAO, the electron antineutrinos 
are generated from the beta decay of 
the fission products. There are 
four major isotopes for nuclear fission in the 
Taishan reactors, $^{235}$U, $^{238}$U, 
$^{239}$Pu, and $^{241}$Pu \cite{JUNO:2020ijm}.
The fission process produces antineutrinos with
energy between 1\,MeV and 10\,MeV.
The production momentum transfer 
$Q^2_p$ is carried by the $W$ boson 
propagator in the conversion of a neutron into
a proton with the emission of the electron and
the electron antineutrino. 
In terms of the neutron ($p_n$), proton ($p_p$),
electron ($p_e$), and antineutrino ($p_\nu$) momentums,
the production momentum transfer is \cite{Giunti:2007ry},
\begin{align}
  Q^2_p
\equiv
  (p_p-p_n)^2
=  
  (p_e+p_{\bar\nu})^2,
\end{align}
with energy-momentum conservation
$p_n = p_p + p_e + p_{\overline \nu}$.
The $Q^2_p$ value ranges from the electron mass 
squared $m^2_e \approx 0.26\,$MeV$^2$ to the square of the mass 
difference between neutron and proton $(m_n-m_p)^2\approx 1.67\,$MeV$^2$ \cite{Giunti:2007ry}.
To be conservative, we take the reference scale
$Q^2_0 = 1$\,MeV$^2$ as a benchmark value to satisfy the
cosmological BBN constraint \cite{Venzor:2020ova} and
remove those running below 1\,MeV. 
Since the $Q^2_p$ range is close to $Q^2_0 = 1$\,MeV$^2$, any variation in $Q^2_p$ will 
result in only small correction to the 
oscillation probability. For simplicity, we 
fix the value of $Q_p^2$ to $(m_n-m_p)^2$ in our study.

The electron 
antineutrino flux will 
be detected at the TAO detector via 
the inverse beta decay 
(IBD) process, $\bar\nu_e + p\rightarrow n + e^+$. 
For this process, an energy threshold of 
1.806\,MeV needs to be considered 
due to the kinematics \cite{Vogel:1999zy}.
The detection momentum transfer is also defined as
the one carried by the $W$ mediator,
\begin{align}
  Q^2_d
\equiv 
  (p_n-p_p)^2. 
\end{align}
In contrast to the production case,
the inverse beta decay reaction 
has a larger phase space and the detection
momentum transfer $Q^2_d$ ranges from $Q^2_{d,{\rm min}}$ 
to $Q^2_{d,{\rm max}}$ as defined \geqn{eq:Q2drange} in the appendix.
For an electron antineutrino energy $E_\nu=10\,$MeV, 
the minimum value $Q^2_{d,{\rm min}}\approx 3.9\times  10^{-2}$\,MeV$^2$
is small but the maximal value $Q^2_{d,{\rm max}}\approx 340.4\,$MeV$^2$
is quite sizable.

\begin{figure}[t!]
\centering
\includegraphics[width=0.48\textwidth]{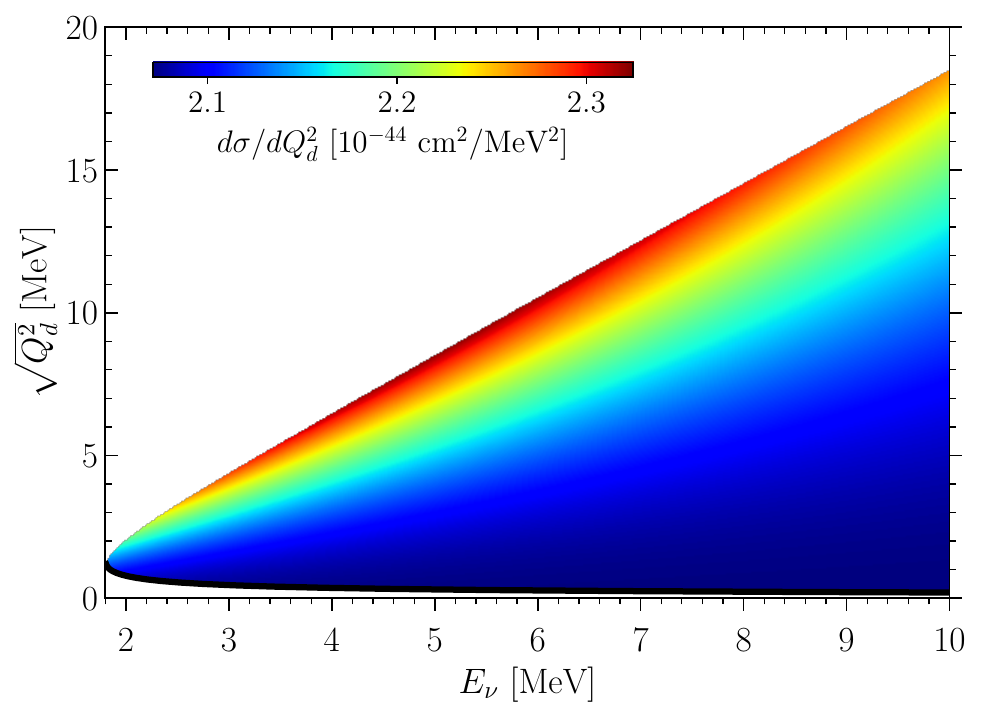}
\caption{The detection differential cross section $d\sigma/dQ^2_d$
on the $E_\nu$--$\sqrt{Q^2_d}$ plane, where the lower black curve corresponds to $\sqrt{Q^2_{d,\rm min}}$.
}
\label{fig:q2-distribution}
\end{figure}

The existence of mismatched momentum transfers
between the production ($Q^2_p \approx 1.67$\,MeV$^2$) and 
detection ($Q^2_d \approx \mathcal O(10^{-2} \sim 10^{2})$\,MeV$^2$) 
processes allows the RG running effect to manifest itself. 
Since the detection momentum transfer $Q^2_d$ spans a
range, its distribution should be taken into account.
The $Q^2_d$ distribution function is described by the
differential cross section, $d\sigma/dQ^2_d$ \cite{Giunti:2007ry}, as follows:
\begin{align}
    \frac{d\sigma}{dQ^2_d}
=&
    \frac{G_F^2\cos^2\theta_C m_p^2}
    {8\pi E_{\nu}^2}
\nonumber 
\\
\times &
    \bigg[ 
    A(Q_d^2)
    -
    B(Q_d^2)
    \frac{s-u}{m_p^2}
    +C(Q_d^2)
    \frac{(s-u)^2}{m_p^4}
    \bigg],
\label{eq:dsigmadQ2}
\end{align}
where $G_F$ is the Fermi constant and $\theta_C$ being the Cabibbo angle.
The Mandelstam variables are $s \equiv (p_\nu + p_p)^2 = m^2_p + 2 m_p E_\nu$
and $u \approx 2 m^2_p + m^2_e - s - t = m^2_p + m^2_e - 2 m_p E_\nu + Q^2_d$
since $Q^2_d \equiv - t$ for a $t$-channel detection process.
The $A$, $B$, and $C$ are form factor,
\begin{subequations}
\begin{align}
  A(Q^2_d)
& \equiv  
  \frac{m_e^2+Q^2_d}{m_p^2}
\Bigg\{
  \left(1+\frac{Q^2_d}{4m_p^2}\right)G_A^2
+ \frac{Q^2_d}{m_p^2}F_1F_2
\nonumber\\
&
- \left(1-\frac{Q^2_d}{4m_p^2}\right)\left(
  F_1^2-\frac{Q^2_d}{4m_p^2}F_2^2
\right)
\nonumber\\
&
- \frac{m_e^2}{4m_p^2} \left[ (F_1+F_2)^2+G_A^2 \right]
    \Bigg\},
\\
  B(Q^2_d)
& \equiv  
  \frac{Q^2_d}{m_p^2}
  G_A (F_1+F_2),
\\
  C(Q^2_d)
& \equiv  
  \frac 1 4
\left( G_A^2+F_1^2+\frac{Q^2_d}{4m_p^2}F_2^2 \right),
\end{align}
\end{subequations}
with
\begin{subequations}
\begin{align}
  G_A
& \equiv 
  \frac{g_A}{\left(1+\frac{Q_d^2}{m_A^2}\right)^2},
\\
  F_2 
& \equiv
  \left(\frac{\mu_p-\mu_n}{\mu_N}-1\right)
  \frac 1 {\left(1+\frac{Q_d^2}{m_V^2}\right)^2 \left(1+\frac{Q_d^2}{4m_p^2}\right)},
\\
  F_1
& \equiv 
  \left(\frac{\mu_p-\mu_n}{\mu_N}\right)
  \frac{1}{\left(1+\frac{Q_d^2}{m_V^2}\right)^2}
- F_2,
\end{align}
\end{subequations}
where $g_A=1.27$ is the axial vector coupling while
$m_A=1.026\,$GeV and $m_V=0.84\,$GeV are the axial
and vector dipole masses, respectively. In addition,
$\mu_p=2.793\,\mu_N$ ($\mu_n=-1.913\,\mu_N$) is the
proton (neutron) magnetic moment.

\gfig{fig:q2-distribution} shows the 
differential cross section
on the $E_\nu$--$\sqrt{Q^2_d}$ plane.
The minimum value $\sqrt{Q^2_{d, \rm min}}$ of $\sqrt{Q^2_d}$ is almost vanishing 
for all neutrino energies while the upper limit $\sqrt{Q^2_{d, \rm max}}$
increases linearly with the neutrino energy $E_\nu$. 
Given $E_\nu$, the differential cross section
increases slowly with $Q^2_d$ in the allowed range.
For the typical energy of reactor neutrinos, the
momentum transfer can reach $\mathcal O(100)$\,MeV$^2$.
With $Q^2_0 = 1$\,MeV$^2$, the RG running
factor $\ln (|Q^2 / Q^2_0|)$ in \geqn{eq:delta-run}
can reach $4 \sim 5$. Consequently, the prefactor of
$\beta_\delta$ in the expanded transition probability
of \geqn{eq:Pee-expanded} is almost 1,
$P_{\rm ee} \approx 1 - \beta^2_\delta$.

\begin{figure}[t!]
\centering
\includegraphics[width=0.48\textwidth]{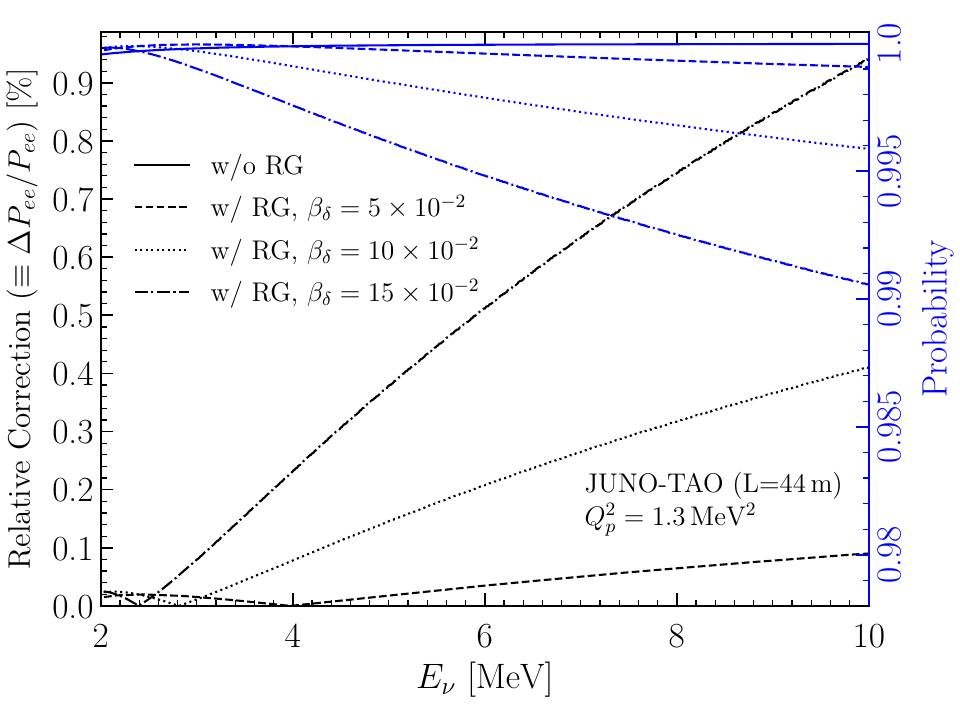}
\caption{
The relative correction of oscillation probability
$\Delta P_{ee}/P_{ee} \equiv |\overline P_{ee}(E_\nu)-P_{ee}(E_\nu)|/P_{ee}(E_\nu)$ 
is shown by the left $y$-axis in black
while the absolute oscillation probabilities
$\overline P_{ee}(E_\nu)$ with (non-solid)
and $P_{ee}(E_\nu)$ without (solid)
RG running effect are
shown by the right $y$-axis in blue.
}
\label{fig:prob}
\end{figure}

To maximize the sensitivity of the RG effect, it
is better to use the two-dimensional distribution 
of neutrino events on the $E_\nu$--$Q^2_d$ plane
with full information of the detection momentum
transfer \cite{Ge:2023azz}. This requires the
directional information of the final-state positron
which is very difficult at TAO \cite{Wei:2020yfs}
since the positron produced from the IBD process
is almost isotropic \cite{Vogel:1999zy}.
So we consider the conservative scenario 
with only the neutrino energy reconstruction. In this case, the 
observed events correspond to the averaged 
oscillation probability $\overline P_{ee} (E_\nu)$
by integrating over the $Q^2_d$ distribution,
\begin{align}
  \overline P_{ee} (E_\nu)
\equiv
  \int P_{ee}(Q^2_{p,d})\frac{1}{\sigma} \frac{d\sigma}{dQ^2_d} d Q^2_d,
\label{eq:aver-P}
\end{align}
where $P_{ee}(Q^2_{p,d})$ is the original oscillation
probability in \geqn{eq:Pee} that contains all information
including the RG running effect. In the conventional
oscillation case, there is no dependence on the momentum 
transfer and hence the oscillation probability 
is labeled by $P_{ee}(E_\nu)$ for comparison
in \gfig{fig:prob}. 
To show the impact of RG running on oscillation, we plot the relative deviation 
$\Delta P_{ee} / P_{ee} \equiv |\overline P_{ee}(E_\nu)-P_{ee}(E_\nu)|/P_{ee}(E_\nu)$ (left $y$-axis) 
together with the oscillation probabilities
$\overline P_{ee}(E_\nu)$ with (non-solid) and
$P_{ee}(E_\nu)$ without (solid) RG running effect
(right $y$-axis).
The conventional oscillation probability  
(blue solid) is very close to 1 as expected. In the
presence of the RG running effect, 
the oscillation probability deviates by up to 1\%
with $\beta_\delta = 0.15$.
The deviation increases with $\beta_\delta$ and
the neutrino energy $E_\nu$ since
a larger neutrino energy leads to a larger
$Q^2_d$ distribution as shown in \gfig{fig:q2-distribution}.

\vspace{2mm}
\textbf{Projected Sensitivity at JUNO-TAO}
--
The Taishan Antineutrino Observatory 
is a satellite experiment of JUNO
and consists of a 2.8 tonne spherical liquid 
scintillator (LS) detector \cite{JUNO:2020ijm}
with a fiducial mass of around 1 tonne.
More than 99.99\% of the antineutrinos 
detected at the TAO detector
come from the Taishan reactor power plant.
The TAO detector is approximately 44\,m away
from one of the two Taishan reactor cores and 217\,m
from the other \cite{JUNO:2024jaw}.
Since the two reactor cores have
the same power, the closer one contributes around 
96\% antineutrinos while the further one 
contributes only 4\%. In the zero-distance limit,
the oscillation probability is a constant as discussed
above. So both reactor fluxes can be simply combined
when studying the RG running effect.

The antineutrino flux from the Taishan reactor 
is modeled as a weighted average of 
isotope fission fraction $f_i$
times the fission spectrum $s_i(E_\nu)\equiv \exp (\sum_{p=1}^6 \alpha_{ip}E_\nu^{p-1})$,
\begin{align}
  \phi(E_\nu)
\equiv 
  \frac{W}{\sum_i f_{i}e_i}
  \sum_i f_{i}s_i(E_\nu),
\label{eq:phi}
\end{align}
where $W$ is the thermal power of reactor, $e_i$ is 
the mean energy released per fission of isotope $i$.
In our study, we take a thermal power of $W=4.6\,$GW
for each reactor core \cite{JUNO:2020ijm} 
and the values of $f_i, e_i, \alpha_{ip}$ 
are taken from \cite{JUNO:2020ijm},
\cite{DayaBay:2016ssb}, and 
\cite{Mueller:2011nm}, respectively. 
Moreover, we divide the energy window of [1.8, 10]\,MeV
with bin size of 50\,keV \cite{JUNO:2020ijm} in our simulation.
The IBD event rate is around 1000 per day in this energy range
\cite{JUNO:2024jaw}.
The expected antineutrino energy spectrum 
at the TAO detector for a fixed distance $L = 44\,$m
can be expressed as,
\begin{align}
  S(E_\nu)
\equiv 
  \frac{N T}{4\pi L^2}
  \phi (E_\nu) \overline P_{ee}(E_\nu)
  \sigma(E_\nu),
\label{eq:N}
\end{align}
where $N$ is the number of free protons in 
the detector target and $T$ is the run time.

In our study, we use GLoBES \cite{Huber:2004ka,Huber:2007ji}
to incorporate the JUNO-TAO
experimental setup and the oscillation 
probability in \geqn{eq:aver-P}. 
Due to the limited energy resolution 
of real detection,
a Gaussian smearing function  
is implemented. We take the 
expected light yield of 4500 
photoelectrons (p.e.) 
per MeV from Silicon
PhotoMultipliers (SiPM) tiles, 
which translates to an energy 
resolution of $1.5\%/\sqrt{E / {\rm MeV}}$ \cite{JUNO:2020ijm} 
where $E$ is the deposit energy, $E \equiv E_\nu - 0.8\,$MeV.

Besides the IBD event, there are three major
backgrounds: the accidental, fast neutron, and 
$^{9}$Li/$^{8}$He backgrounds \cite{Basto-Gonzalez:2021aus}.
Their spectra are extracted from \cite{JUNO:2024jaw}.
In the [1.8, 10]\,MeV energy window,
the background event rate is around (170, 67, 51) per day for the 
(accidental, fast neutron, $^{9}$Li/$^{8}$He)
component, respectively.

\begin{figure}[t!]
\centering
\includegraphics[width=0.48\textwidth]{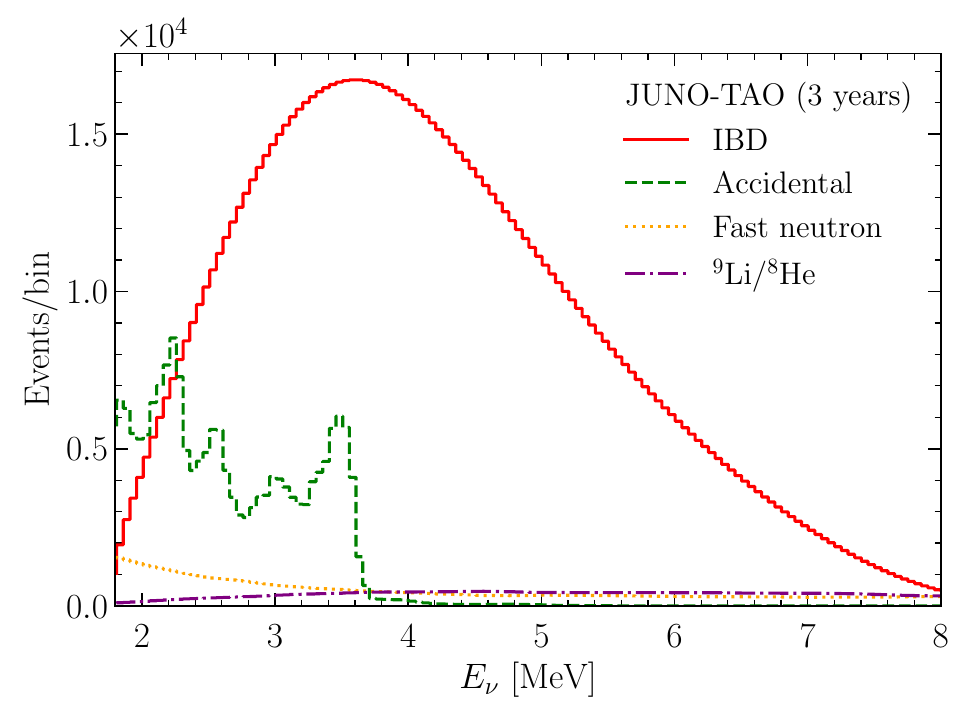}
\caption{
The energy spectra for IBD (red), accidental (dashed green),
fast neutron (dotted orange), and $^{9}$Li/$^{8}$He (dash-dotted purple) events without oscillation. 
The background spectra are extracted from \cite{JUNO:2024jaw}.
}
\label{fig:event}
\end{figure}

The event distributions in the TAO detector
without any oscillation or transition 
are shown in \gfig{fig:event}.
The IBD event spectrum has a bell
shape with the maximum around $(3.5 \sim 4.0)$\,MeV. 
For background, the accidental one (green curve) 
distributes mostly below 3.6\,MeV and drops
fast as energy increases. In comparison, the 
fast neutron (orange) and 
$^{9}$Li/$^{8}$He (purple) background 
components are flat. Since the 
RG running effect increases with energy
as shown in \gfig{fig:prob}, 
the low background at $E_\nu>3.5\,$MeV
makes JUNO-TAO particularly sensitive to non-zero values of $\beta_\delta$.

To quantify the JUNO-TAO sensitivity to
the RG running parameter $\beta_\delta$, we
adopt the following $\chi^2$ function \cite{JUNO:2020ijm},
\begin{align}
  \chi^2
& \equiv 
  \sum_i^{\rm bins}
  \left(\frac{N_{i}^{\rm true} - N_{i}^{\rm test}
      }{\sqrt{N_{i}^{\rm true}}} \right)^2
  + 
  \chi^2_{\rm para}
\nonumber\\
& \
  +\left(\frac{a_{R}}{\sigma_{a_{ R}}}\right)^2
  +\left(\frac{b_{R}}{\sigma_{b_{R}}}\right)^2  
  +\left(\frac{a_{A}}{\sigma_{a_{ A}}}\right)^2
  +\left(\frac{b_{A}}{\sigma_{b_{ A}}}\right)^2
\nonumber\\
& \
  +\left(\frac{a_{L}}{\sigma_{a_{ L}}}\right)^2  
  +\left(\frac{b_{L}}{\sigma_{b_{ L}}}\right)^2
  +\left(\frac{b_{F}}{\sigma_{b_{ F}}}\right)^2,
\label{eq:chi2}
\end{align}
where the first term on the right-hand side
corresponds to the
Gaussian approximation for the binned
events. 
In the first term, $N_{i}^{\rm true}\equiv N_{i}^{\rm sig, true}+N_{i}^{\rm bkg, true}$ 
is the total true event number within the $i$-th energy bin
without oscillation. Its counterpart $N_{i}^{\rm test}$
is defined as \cite{JUNO:2020ijm},
\begin{align}
\hspace{-3mm}
   N_{i}^{\rm test}
& \equiv 
  (1+a_{R})R_{i} 
  +
  (1+a_{A})
  A_{i} 
  + (1+a_{L})L_{i}
  + F_{i}
\nonumber \\
& +
  (b_{R} R_{i}+ b_{A} A_{i}+ b_{F} F_{i}+b_{L} L_{i}) 
  \frac{E_i' - \bar E'}{E'_{\rm max}-E'_{\rm min}},
\label{eq:N-test}
\end{align}
where the IBD signal event number $R_{i}$ contains
the test RG parameter $\beta_\delta$. The remaining
$(A_{i}, F_{i}, L_{i})$ are the event numbers for
the (accidental, fast neutron, $^{9}$Li/$^{8}$He) background
components in the $i$-th energy bin, respectively.

As shown in \geqn{eq:Pee}, only $\theta_{13}$ and
the Dirac CP phase difference $\Delta \delta_D$
affect the oscillation probability. So we include
a Gaussian prior 
$\sin^2 \theta_{13}/10^{-2} = 2.200^{+0.069}_{-0.062}$
\cite{deSalas:2020pgw} in the $\chi^2_{\rm para}$ function.
Since the $\theta_{13}$ mixing angle has been measured
very precisely by mainly Daya Bay \cite{DayaBay:2022orm}
as well as RENO \cite{Shin:2020mue} and Double CHOOZ
\cite{Soldin:2024fgt}, it actually does not expect to
spoil the sensitivity on the RG parameter $\beta_\delta$.

The remaining terms describe the systematic uncertainties.
For each signal and background event number,
we assign overall scaling ($a_i$) and
tilt ($b_i$) nuisance parameters for the
normalization and shape uncertainties, respectively.
First, each component has a
nuisance parameter $a_X$ ($X\in\left\{{R, A, L}\right\}$)
for the normalization uncertainty, $\sigma_{a_R} = 
10\%$ (IBD signal), $\sigma_{a_A} = 1\%$ (accidental), and
$\sigma_{a_L} = 20\%$ ($^{9}$Li/$^{8}$He),
respectively \cite{JUNO:2024jaw}.
Note that the fast neutron component has no normalization
uncertainty since it can be very precisely
measured using the reactor-off data \cite{JUNO:2024jaw}.

Second, we follow GLoBES \cite{Huber:2004ka,Huber:2007ji} to take a
linear tilt to parametrize
the overall shape uncertainty. Over the reconstructed
energy range $[E'_{\rm min}, E'_{\rm max}]
\equiv$ [1.8, 10]\,MeV,
the spectrum would tilt around the reference
energy point $\bar E'\equiv (E'_{\rm max}+E'_{\rm min})/2$.
While the overall normalization uncertainty is energy
independent, the variation due to shape uncertainty
increases with the reconstructed neutrino energy
$E'$ and can fake the RG running effect.
For the IBD signal,
a 1\% uncertainty can be estimated from the binned
uncertainties in Fig.6 of \cite{JUNO:2024jaw},
including a 0.6\% statistical contribution and a
0.8\% systematical one.
\begin{figure}[t!]
\centering
\includegraphics[width=0.48\textwidth]{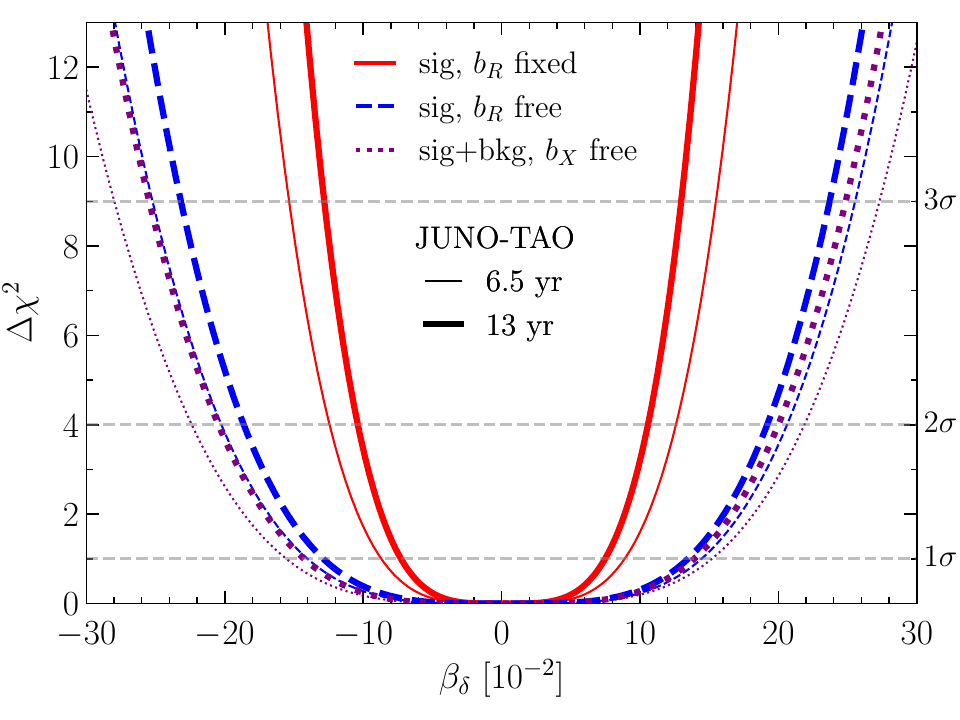}
\caption{
The expected sensitivities ($\Delta \chi^2$) of the RG running
parameter $\beta_\delta$ at JUNO-TAO.}
\label{fig:result}
\end{figure}

The JUNO-TAO sensitivity to $\beta_\delta$ presented 
in \gfig{fig:result} is obtained after marginalizing
$\chi^2$ over all the nuisance parameters and $\theta_{13}$.
The solid lines are the results obtained for the IBD signal
only without considering
the backgrounds or the spectrum tilt.
Currently, the JUNO-TAO detector is designed to run for
6.5 years (thin) and we assume a possible extension 
to 13 years (thick) for comparison. With larger statistics,
the $\beta_\delta$ sensitivity receives quite sizable
improvement due to reduction of the statistical fluctuation.
With 2.4 million events for 6.5 years to be expected at
JUNO-TAO \cite{JUNO:2024jaw}, the pure statistical fluctuation
$\sqrt{N^{\rm true}_i}$ of event numbers can the spectrum
to tilt up to 0.6\%.
As estimated above, the zero-distance transition probability
scales with $\beta_\delta$ roughly as
$P_{\rm ee} \approx 1 - \beta^2_\delta$ in the middle
of the energy range. The 0.6\% spectrum shape variation would
translate to a 7.7\% uncertainty of $\beta_\delta$ which
is consistent with the solid curve in \gfig{fig:result}.

Setting free the tilt parameter $b_R$ for the IBD signal
can significantly deteriorate the sensitivity, which is shown
as the blue dashed curves. This clearly shows that the
spectrum tilt uncertainty is very similar to the RG running
effect that is shown in \gfig{fig:prob} and consequently
becomes a key factor for the $\beta_\delta$ sensitivity.
It is desirable to explore the possibility of experimentally
further suppressing the tilt uncertainty.

In comparison, the backgrounds would not make that large trouble. As shown with the purple dotted lines in
\gfig{fig:result}, the sensitivity becomes only slightly
worse when the backgrounds and their spectrum uncertainties,
including both normalization and tilt, are switched on.
This is because the RG running effect grows with the
neutrino energy $E_\nu$ but the backgrounds mainly
appears in the low-energy region as demonstrated in
\gfig{fig:event}. The influence of backgrounds is naturally
suppressed in the search of the RG running effect.

\vspace{2mm}
\textbf{Conclusion}
--
Reactor neutrino oscillation experiments are usually
envisioned as being able to measure the $\theta_{13}$ and
$\theta_{12}$ mixing angles, but the Dirac CP phase $\delta_D$
cannot appear in the relevant oscillation probability $P_{ee}$.
However, we point out that the upcoming JUNO-TAO experiment
can provide independent sensitivities to the RG
running effect of the Dirac CP phase which appears
as the deviation in $\delta_D$. The projected sensitivity
on $\beta_\delta$ can reach around 10\%. In other words,
it is possible for reactor experiments to probe
CP-related new physics. With much smaller momentum
transfer, around only 1\,MeV for the beta decay
branches, reactor experiments have the advantage
of probing new physics at much lower energy.
The sensitivity should further improve if the
final-state positron direction can be reconstructed
to obtain the full information of momentum transfer
in the detection process and the spectrum tilt
uncertainty should also be further improved.

\vspace{2mm}
\textbf{Data Availability}
--
No data were created or analyzed in this study.

\vspace{2mm}
\textbf{Acknowledgements}
--
The authors are grateful to Liang-Jian Wen,
Yi-Chen Li, and Han Zhang for kind help.
This work is supported by the National Natural Science
Foundation of China (12425506, 12375101, 12090060 and
12090064) and the SJTU Double First
Class start-up fund (WF220442604).
SFG is also an affiliate member of Kavli IPMU, University of Tokyo.
 PSP is also supported by the Grant-in-Aid for Innovative Areas No. 19H05810.

\appendix
\section{Appendix: Detection Momentum Transfer}

\renewcommand{\theequation}{A\arabic{equation}}
\setcounter{equation}{0}

To find the $Q^2_d$ range of the detection process
$\bar \nu_e + p \rightarrow e^+ + n$,
which can be parametrized as a general
two-body scattering process,
we work in the center-of-mass (CM) frame. 
For clarity, we use asterisk 
$^\ast$ to label the kinematic variables 
in the CM frame,
\begin{align}
 (E^\ast_\nu, {\bf p}^\ast_\nu)+
 (E^\ast_p, {\bf p}^\ast_p)
 \longrightarrow
 (E^\ast_e, {\bf p}^\ast_e)+
 (E^\ast_n, {\bf p}^\ast_n).
\end{align}
For a two-body system in the CM frame, their
energy and momentum are functions of the total
energy \cite{PDG22-kin},
\begin{subequations}
\begin{align}
\hspace{-2mm}
  E^\ast_\nu
& = \frac{s - m_p^2}{2\sqrt{s}},
\quad 
 E_p^\ast =\frac{s+m_p^2}{2\sqrt{s}},
\\
\hspace{-2mm}
    E_e^\ast
& = \frac{s+m_e^2-m_n^2}{2\sqrt{s}},
\quad 
    E_n^\ast =\frac{s+m_n^2-m_e^2}{2\sqrt{s}},
\\
\hspace{-2mm}
  {\bf p}_\nu^\ast
& ={\bf p}_p^\ast 
= \frac{(s-m_p^2)}{2\sqrt{s}},
\\
\hspace{-2mm}
  {\bf p}_e^\ast
& = {\bf p}_n^\ast 
= \frac{\sqrt{[s-(m_e-m_n)^2][s-(m_e+m_n)^2]}}{2\sqrt{s}},
\label{eq:com_kinematics}
\end{align}
\end{subequations}
with the Lorentz-invariant CM energy
$s\equiv (p_\nu^\ast +p_p^\ast)^2 = m^2_p + 2 m_p E_\nu$. 

The detection momentum transfer $Q^2_d \equiv 
- (p_\nu^\ast - p_e^\ast)^2$  in the CM frame,
\begin{align}
\hspace{-3mm}
  Q^2_d
= 
    \left(
    |{\bf p}_\nu^\ast|^2+|{\bf p}_e^\ast|^2-2|{\bf p}_\nu^\ast||{\bf p}_e^\ast|\cos\theta^\ast
    \right)
    -(
    E^\ast_\nu-E^\ast_e)^2,
\nonumber
\end{align}
then depends on the angle $\theta^\ast$ between 
${\bf p}_\nu^\ast$ and ${\bf p}_e^\ast$.
In the CM frame the scattering angle 
$\theta^\ast$ is arbitrary, 
and the minimum value of $Q^2_d$ occurs when the
initial neutrino momentum ${\bf p}_\nu^\ast$
is anti-parallel to the final-state positron
momentum ${\bf p}_e^\ast$,
\begin{subequations}
\begin{align}
    Q^2_{d, {\rm min}}
=
    \left(
    |{\bf p}_\nu^\ast|-|{\bf p}_e^\ast|
    \right)^2
    -(
    E^\ast_\nu - E^\ast_e)^2,
    \label{eq:qmin}
\end{align}
while the maximum 
value of $Q^2_d$ occurs when ${\bf p}_\nu^\ast$
has the same direction as ${\bf p}_e^\ast$,
\begin{align}
    Q^2_{d,{\rm max}}
=
    \left(
    |{\bf p}_\nu^\ast|+|{\bf p}_e^\ast|
    \right)^2
    -(
    E^\ast_\nu - E^\ast_e)^2.
    \label{eq:qmax}
\end{align}
\label{eq:Q2drange}
\end{subequations}
One may see that there is cancellation between
the momentum and energy terms in $Q^2_{d, \rm min}$
which indicates that $Q^2_{d, \rm min}$ should
be quite small while $Q^2_{d, \rm max}$ can be
large.

\addcontentsline{toc}{section}{References}
\bibliographystyle{plain}

\end{document}